\documentclass[multphys,vecphys]{svmult}
\usepackage{makeidx}         
\usepackage{graphicx}        
\usepackage{multicol}        
\usepackage[bottom]{footmisc}
\makeindex             

\begin{document}

\title*{Photometric and kinematical study of nearby groups of galaxies around IC 65 and NGC 6962}
\titlerunning{Nearby groups of galaxies around IC 65 and NGC 6962} 
\author{J. Vennik\inst{}\and E. Tago\inst{}}
\institute{Tartu Observatory, 61602 T\~oravere, Tartumaa, Estonia \\ 
\texttt{vennik@aai.ee, erik@aai.ee}}
\maketitle

\section{The IC 65 group}
\label{sec:2}

The IC 65 group (z = 0.0089) of four late type galaxies - IC 65, UGC 608, UGC 622, and PGC 138291 - 
has been studied earlier in the 21 cm HI line 
by van Moorsel (1983, A\&AS, 54, 1), who found disturbed 
HI envelopes of bright group members, and detected a new HI-rich LSB galaxy. 
We have searched for new dwarf member candidates of this group 
on the calibrated DSS 2 blue and red frames by means of SExtractor software.  
Dwarf galaxies were selected using the 
surface brightness (SB), light concentration, colour and morphological criteria. 
As a result, we have selected four LSB irregular blue galaxies, which 
fit the empirical SB - magnitude  
relation for dwarf galaxies (Ferguson \& Binggeli 1994, A\&ARv, 6, 67).
A detailed surface photometry of selected galaxies has been carried out on the $B, R$ and $I$ CCD 
frames obtained at Calar Alto (Vennik \& Hopp 2004, astro-ph/0409632).\\
{\it Results for the IC 65 group:}
The group consists of two dense subgroups 
with a projected separation of $\sim$ 220 kpc 
($H_0 = 75$ km s$^{-1}$ Mpc$^{-1}$ is assumed throughout). 
The bright group members are rich in HI, 
their outer HI isophotes generally appear disturbed.  
Newly found probable dwarf companions are of 
irregular and/or of the head-tail shape with blue star-forming knots. 
We conclude that the IC 65 group of galaxies is 
dynamically young and possibly at the stage of its first collapse with some evidence of 
(tidal) interactions between its members.
\section{The NGC 6962 group}
\label{sec:3}
The NGC 6962 group (z = 0.014) is a rich assembly of galaxies, which consists of up to 
28 galaxies with concordant redshifts, listed in the NED within 2.6$^o$ 
($\sim$ 2.6 Mpc) around the principal galaxy. We have extracted the list of 
the dwarf galaxy candidates for the group from the {\it PhotoObjAll} catalog of the SDSS DR4, 
using the SB, light concentration and isophotal diameter selection criteria. 
The pre-selected candidates have visually been inspected on the SDSS $g, r$ and $i$ 
frames and the final membership probability (rated 1 - 3) has been assigned on the 
morphological, SB and colour grounds.  
We compared the angular correlation functions (CFs) of true members (reference sample) 
and newly selected galaxies of different priority classes.  
The CF of the highest probability (rated 1) candidates could have been reproduced  
by a mix of 55\% of reference sample and 45\% of the randomly distributed objects.  
The rated 2 and 3 ensembles are more severly contaminated by background galaxies.
\begin{figure}
\resizebox{0.46\textwidth}
{!}{\includegraphics*{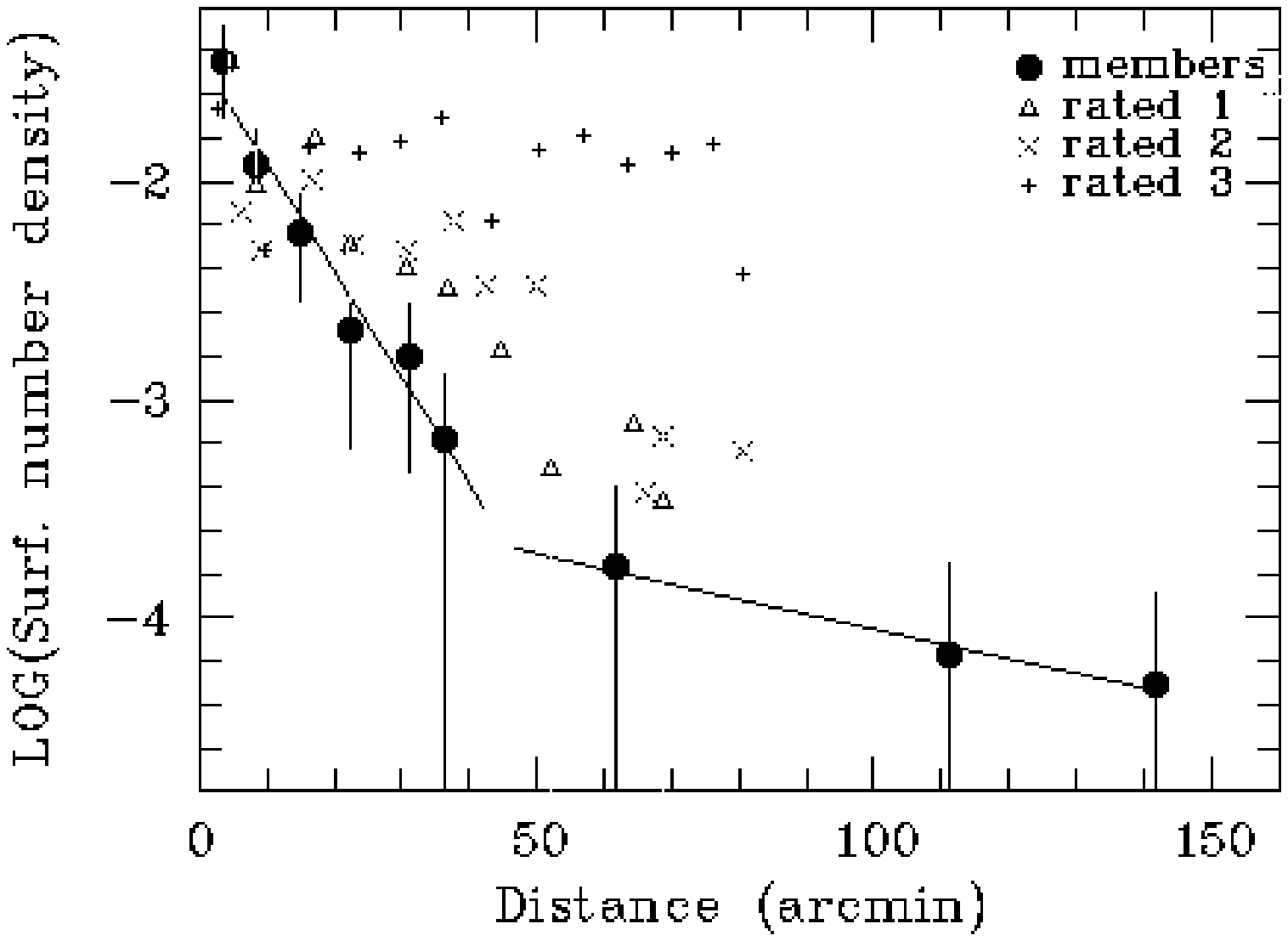}}
\hspace{3mm} 
\resizebox{0.5\textwidth}
{!}{\includegraphics*{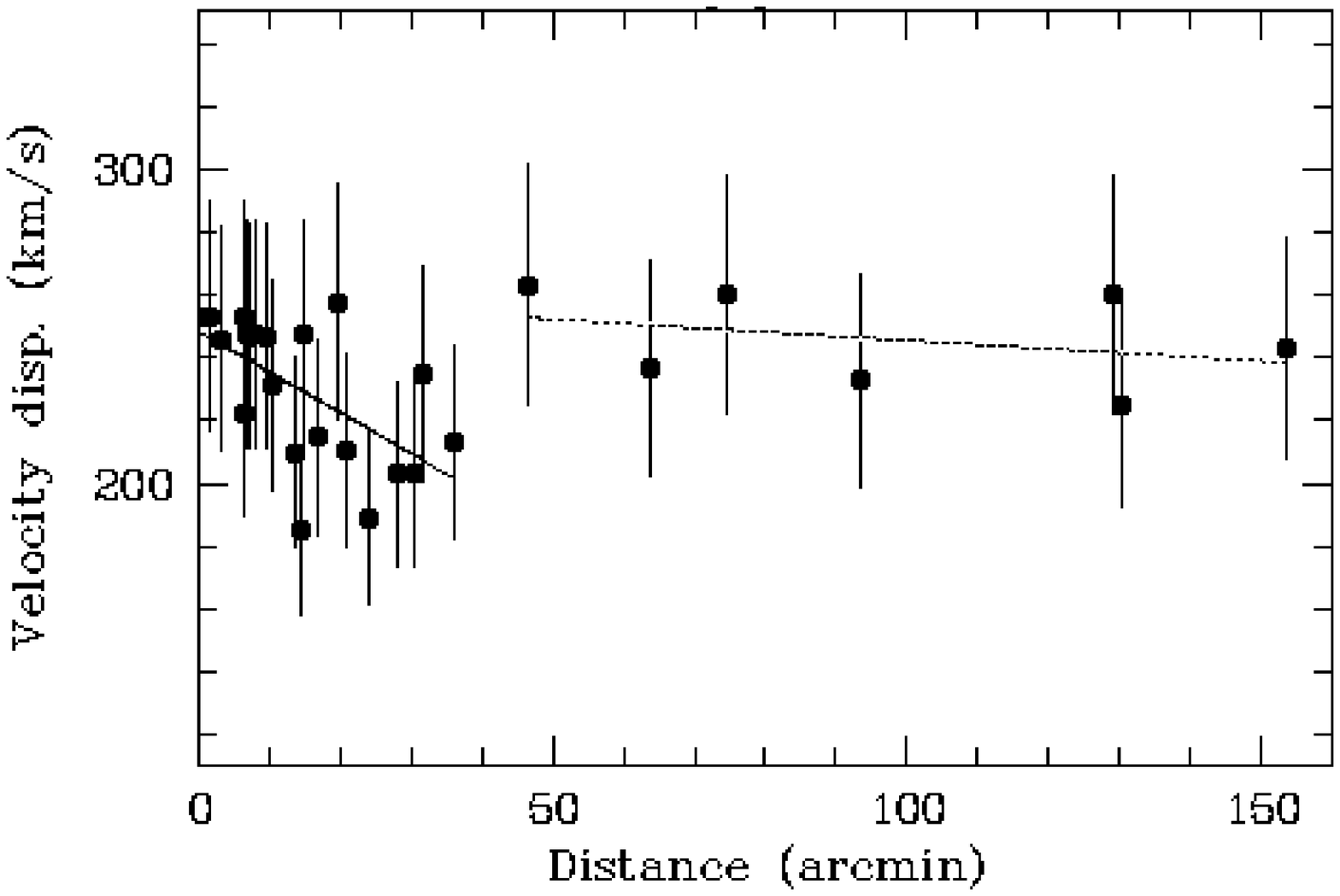}}
\caption{
Number density ({\it left}) and velocity dispersion ({\it right}) profiles of true members  
($\bullet$) and dwarf member candidates (rated 1, 2, 3) in the NGC 6962 group.}
\end{figure}
The group displays a dense core + a sparse halo structure. The number density profile  
(Fig 1, left) and the radial velocity dispersion ($\sigma_v$) profile  
of a moving window of 9 galaxies (Fig 1, right) both display a break at 
$\sim 45'$ (730 kpc), which could be interpreted as a core radius  
or the rebound radius of the group. Within the rebound radius the $\sigma_v$ 
profile is marginally dropping, indicating some dynamical evolution in the core region.\\ 
{\it Results for the NGC 6962 group:} 
The group has a well-defined partially relaxed core + halo (or infall region) structure.
Further evidence of an evolved group is provided by clear morphological/spectral segregation, 
with the core being populated mainly  
by passive S0 and E+ galaxies, except the SXab type principal galaxy.
Non-detection of diffuse X-ray emission by RASS indicates that the (supposed) intragroup gas  
\begin{table}
\centering
\caption{Parameters of the IC 65 and NGC 6962 (core) groups of galaxies}
\label{tab:1}       
\begin{tabular}{c|c|c|c|c|c|c|c|c|c} 
\hline\noalign{\smallskip}
Group & n$_{gal}$ & $L_T$ & $R_{harm}$ & $<R_{ij}>$ & $<v_0>$ & $\sigma_v$ & ${\cal M}_{vt}$ 
& ${\cal M}_{vt}/L_T$ & $t_{cr} H_0$ \\
& & $10^{10}L_{\odot}$ & kpc & kpc  & km s$^{-1}$& km s$^{-1}$& $10^{12} {\cal M}_{\odot}$& \\
\noalign{\smallskip}\hline\noalign{\smallskip}
IC 65 & 5 & 5.8 ($B$) & 135 & 192 & 2890 & 77 & 2.2 & 38 & 0.23 \\
NGC 6962 & 23 & 13.4 ($g$) & 237 & 456 & 4200 & 238 & 32.6 & 242 & 0.13 \\
\noalign{\smallskip}\hline
\end{tabular}
\end{table}
should be in a relatively cold stage.
{\it Acknowledgments.} The reserach was supported by the Estonian Science Foundation grants 6104 and 6106.
This study has made use of the NASA/IPAC Extragalactic Database (NED), the STScI Digitized Sky Survey 
(DSS), and the Sloan Digital Sky Survey (SDSS).
\end{document}